\documentclass[12pt,onecolumn]{revtex4-1}
\usepackage[pdftex]{graphicx}
\usepackage{url}
\usepackage{color}
\usepackage{soul}
\newcommand{\registered}{$^{\tiny{\textregistered}}~$}
\newcommand{\PRL}{Physical Review Letters}
\newcommand{\JHEP}{Journal of High Energy Physics}
\newcommand{\PR}{Physical Review}
\newcommand{\NIM}{Nuclear Instruments and Methods in Physics}

\begin{document}

\title{Development of an FPGA-based realtime DAQ system for axion haloscope experiments}

\author{MyeongJae Lee}
\email{myeongjaelee@ibs.re.kr}
\affiliation{Center for Axion and Precision Physics research, Institute for Basic Science, Daejeon 34051, Republic of Korea}
\author{ByeongRok Ko}
\affiliation{Center for Axion and Precision Physics research, Institute for Basic Science, Daejeon 34051, Republic of Korea}
\author{Saebyeok Ahn}
\affiliation{Department of Physics, Korea Advanced Institute of Science and Technology, Daejeon 34141, Republic of Korea}

%

\begin{abstract}
A real-time Data Acquisition (DAQ) system for the CULTASK axion haloscope experiment was constructed and tested. 
The CULTASK is an experiment to search for cosmic axions using resonant cavities, 
to detect photons from axion conversion through the inverse Primakoff effect in a few  GHz frequency range in a very high magnetic field 
and at an ultra low temperature. 
The constructed DAQ system utilizes a Field Programmable Gate Array (FPGA) for data processing and Fast Fourier Transformation.
This design  along with a custom Ethernet packet designed for real-time  data transfer enables 100\% DAQ efficiency, 
which is the key feature compared with a commercial spectrum analyzer. 
This DAQ system is optimally designed for RF signal detection in the axion experiment, 
with 100\,Hz frequency resolution and 500\,kHz analysis window. 
The noise level of the DAQ system averaged over 100,000 measurements is  around -111.7\,dBm. 
From a pseudo-data analysis, an improvement of the signal-to-noise ratio due 
to repeating and averaging the measurements using this real-time DAQ system was confirmed. 
\end{abstract}

%

%
%
%

\maketitle

\section{Introduction}
\label{sec:introduction}
Axions\,\cite{ref:axion1,ref:axion2,ref:axion3,ref:axion4}, candidates for the dark matter in the Universe, 
can be detected 
by their coupling with photons (magnetic field) and  conversion to another photon 
(the inverse Primakov effect). The frequency of the converted photon corresponds to  
Radio Frequency (RF) wavelengths, ranging from a few hundreds MHz to a few tens of GHz, 
when the expected axion mass is a few $\mu$eV. 
Accordingly, an axion  can be detected by using a radio antenna with a resonant cavity in a magnetic field, and RF receiver electronics.
The signal appears as a peak in the frequency domain, and  the width of this signal (a few kHz)
depends on the velocity distribution of the axion in the universe. 
CULTASK (CAPP Ultra-Low Temperature Axion Search in Korea)\,\cite{ref:cultask1,ref:cultask2,ref:cultask3} 
is one of such 
axion haloscope experiments  at Center for Axion and Precision Physics research (CAPP). 

\begin{figure}[b]
\begin{center}
\includegraphics[width=0.9\textwidth]{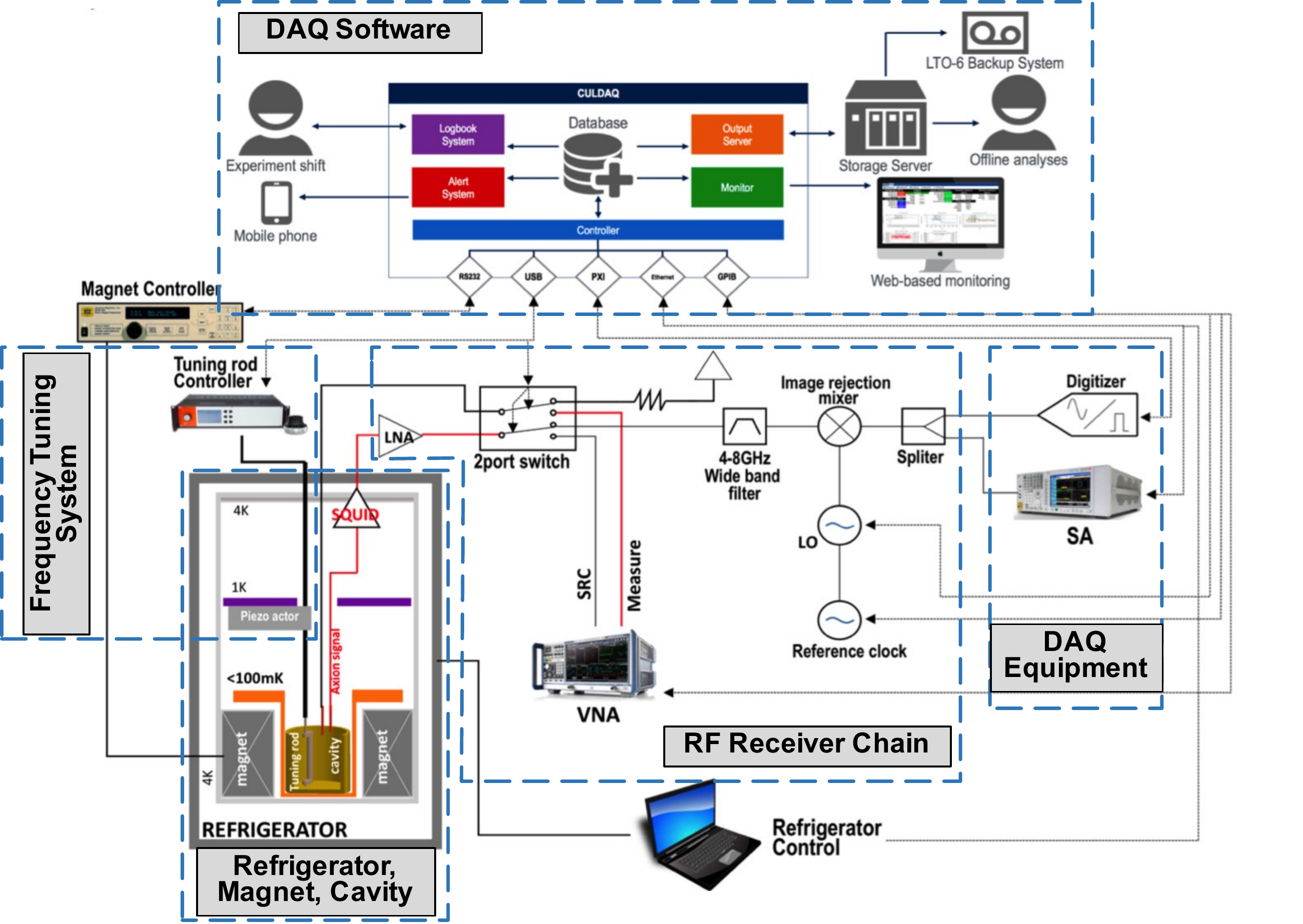}
\caption{System diagram of the experimental setup of the CULTASK experiment. See the main text for detailed explanation. }
\label{fig:systemdiagram}
\end{center}
\end{figure}

Figure \ref{fig:systemdiagram} shows the typical experimental setup of the CULTASK experiment.
In order to scan a frequency range of the axion conversion, the resonant frequency of the cavity is tuned
by using a Frequency Tuning System, so that the resonant frequency matches with the axion conversion frequency being measured. 
The measured signal is amplified by a series of cryogenic amplifiers, 
followed by a super-heterodyne RF Receiver Chain, as shown in the figure. 
In the super-heterodyne receiver design, the RF signal is down-converted
to a fixed Intermediate Frequency (IF), by changing the Local Oscillator (LO) frequency,  
so that  Data AcQuisition (DAQ) Equipment can measure the same frequency range every time. 

The power of the photon converted from the axion ($P$) can be written:
\begin{equation}
P \propto g_{a\gamma\gamma} V B^2 \rho_a \frac{1}{\epsilon} C_{nl}\,{\rm min}(Q_{nl},Q_a) \frac{1}{m_a}~,
\label{eq:axionpower}
\end{equation}
where $g_{a\gamma\gamma}$ is a coupling coefficient between the axion and the photon, 
$V$ is the volume of the resonant cavity,
$B$ is the magnetic field strength, 
$\rho_a$ is the energy density of the halo axion, 
$\epsilon$ is the dielectric constant of vacuum,
$C_{nl}$ is a resonant mode-dependent form factor of the cavity, 
$Q_{nl}$ is the quality factor of the cavity, 
$Q_a$ is the quality factor of the halo axion, and 
$m_a$ is the mass of the axion\,\cite{ref:sikivie, ref:admx}.
The signal power 
depends linearly on the volume of the resonant cavity, 
the square of the magnetic field, and the quality factor of the resonant cavity. 
Combining Eq.\,\ref{eq:axionpower} with the radiometer equation described in Eq.\,\ref{eq:radiometer} in Section\,\ref{sec:test:pseudodata},
the data scan time between frequency $f_1$ and $f_2$ depends on system noise, and 
${\rm scan~time} \sim (T_S)^2 (1/f_1 - 1/f_2 )$,
where $T_S$ = (System noise temperature) = $T_C$(cavity temperature) + $T_A$(amplifier noise temperature).
Therefore, developing a large volume cavity, increasing the magnetic field, 
improving the quality factor of the resonant cavity,
and decreasing the system temperature improve the axion search sensitivity.
In addition to  cavity system development,  signal processing is also important in order not to 
lose the very weak signal from the cavity. A low noise preamplifier located inside the refrigerator is critical
to amplify the cavity signal to a measurable power level with minimum additional noise. 
Signal processing from room temperature  down to the DAQ system is required to
add least amount of additional noise or spurious artifact signal. The axion data, 
i.e., the noise spectrum or its excess  due to the 
addition of the axion signal, is recorded by a spectrum analyzer.

While experimental sensitivity is mostly determined by the noise temperature, the statistics of the data 
are also critical, to achieve better sensitivity within a limited experimental running time. 
High performance DAQ equipment, such as a vector spectrum analyzer or a Fast Analog-Digital Converter (FADC) system  
may enable data taking with small data loss. 

The spectrum analyzer measures the signal using an Analog-Digital Converter (ADC) after analog signal processing. 
The ADC data is then Fourier-transformed to produce a power spectrum.  
Most modern vector spectrum analyzers are  robust and proven, and 
are capable of measuring very small signals. However,
this requires scanning and averaging the spectrum over the frequency range of the measurement. 
In fact, the online process of  analog signal processing and digital manipulation requires a lot more time than 
the actual signal measurement time, which is one reason for the inefficient data taking 
in an axion haloscope experiment.  
This inefficiency also increases when  the Resolution BandWidth (RBW) of the spectrum analyzer is reduced.  
A  so-called ``real-time'' spectrum analyzer is capable of taking the spectrum data without inefficiency, 
but  the RBW  is limited in many cases, and only the averaged spectrum over repeated measurements is  
available as a final result, resulting in the loss of a lot more experimental data which could  
provide a clue in the axion search.  

In some cases, an FADC based DAQ system combined with an offline Fast Fourier Transform (FFT) analysis 
can be used to record the data in the time-domain, instead of using a spectrum 
analyzer and recording the power spectrum in the frequency-domain. 
Axion experimental data  is not conventional compared to various high energy physics experiments which records
the waveform of a charge or voltage signal when triggered. In the axion experiments, the RF signal is
continuously monitored or recorded without a trigger and the resulting power spectrum becomes the
final data. Because of  this difference, an FADC with pipelined post-processing including FFT 
is required in the DAQ system. 
The benefit of an FADC system is that it enables 
much smaller RBW analysis when the sampling rate is very high. 
The critical point in this case is the huge data size transferred from the DAQ to the offline 
computing system for FFT, due to the high sampling rate. For example, 
the data rate  for a  1\,GHz sampling rate FADC is around 10\,Gbps (also considering serial data encoding), 
which is marginal even using 10\,GbE Ethernet. 
Huge data storage with fast writing speed is also required.

Therefore, the  spectrum analyzers or FADC DAQ equipment on the market are not optimal DAQ equipment 
for axion experiment data taking. 
Here, a DAQ system with online FFT based on an FPGA (Field Programmable Gate Array) was developed to
mitigate the major issues of 
(1) short live time (i.e.,  signal measurement time) compared to the post-processing time, 
which results in low DAQ efficiency, and
(2) high data throughput and big data size when using FADC with a few hundreds of MHz to GHz sampling rate.
The FPGA is the optimal device to implement the pipelined post-processing of data. 
This highly efficient DAQ system, optimized  for the CULTASK axion experiment, 
will enable a new target sensitivity limit to be reached, with
limited experiment time and budget,
eventually opening up new physics capability.

This paper describes the development and  performance 
of the new DAQ system, which runs in real-time without  loss of data
due to signal processing procedure. 
The system design of both the hardware and FPGA firmware are described in Sections\,\ref{sec:hardware} and \ref{sec:firmware}. 
The performance of the system is described in Section\,\ref{sec:test}. 
The developed FPGA-based custom real-time DAQ system will be simply  called the {\it FPGA DAQ system} hereafter.

\section{Design of the FPGA DAQ system hardware for axion experiments}
\label{sec:hardware}

\subsection{Overall design of the FPGA DAQ hardware}
\label{sec:hardware:overall}

The FPGA DAQ system was designed and constructed 
to increase  data taking efficiency to 100\%, 
and eventually to reach better sensitivity with limited DAQ time. 
The requirements of the FPGA DAQ system are :
\begin{itemize}
\item[-]  Analysis window of 500\,kHz
\item[-]  Frequency width of 100\,Hz
\item[-]  Averaged noise level less than -100\,dBm.
\end{itemize}

\begin{figure}
\begin{center}
\includegraphics[width=0.9\textwidth]{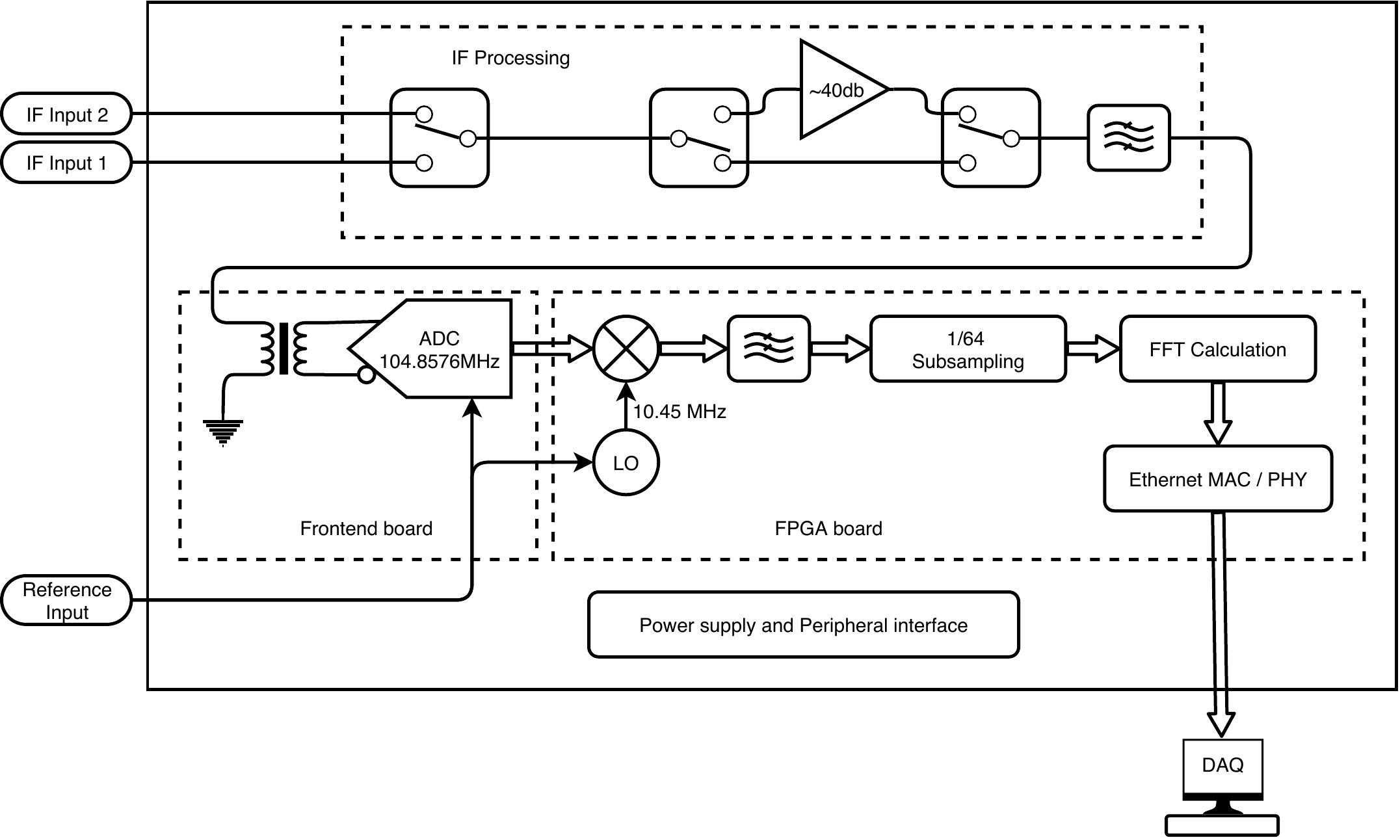}
\caption{The system diagram of the FPGA DAQ system. The solid line and box arrow represent the analog 
and digital signal flows, respectively.}
\label{fig:fpga_digitizer}
\end{center}
\end{figure}

Figure \ref{fig:fpga_digitizer} shows the block diagram  of the FPGA DAQ system, 
consisting of
three core subsections, an RF processing path, frontend FADC board, and FPGA board. 
Because  the frequency of the axion signal in CULTASK experiments is a few GHz, 
the signal is first down-converted 
using an external super-heterodyne RF receiver chain, as shown in Fig. \ref{fig:systemdiagram}. 
The IF of the down-conversion 
of this external room temperature signal processing chain is set to 10.7\,MHz, 
which is a conventional value for an  FM radio receiver. 
After the pre-amplification or channel selection of the 
IF signal
from the axion experiment through the 
IF processing path, 
the signal  is digitized in the frontend board and sent to the FPGA board.  
The subsequent down-conversion directly to baseband 
and band-pass filtering is purely digital inside the FPGA logic,
therefore they are immune to additional analog noise\footnote{The quantization noise, i.e., digital noise may be 
negligible compared to the dynamic range of the ADC.}. 
The FFT process with sub-sampled data is performed in real-time inside the FPGA, therefore, 
there is no dead time in obtaining and transmitting the FFT data.
The system  also features  a processor core unit which cooperates 
with the system interfaces and monitor. 

\begin{figure}
\begin{center}
\includegraphics[width=0.8\textwidth]{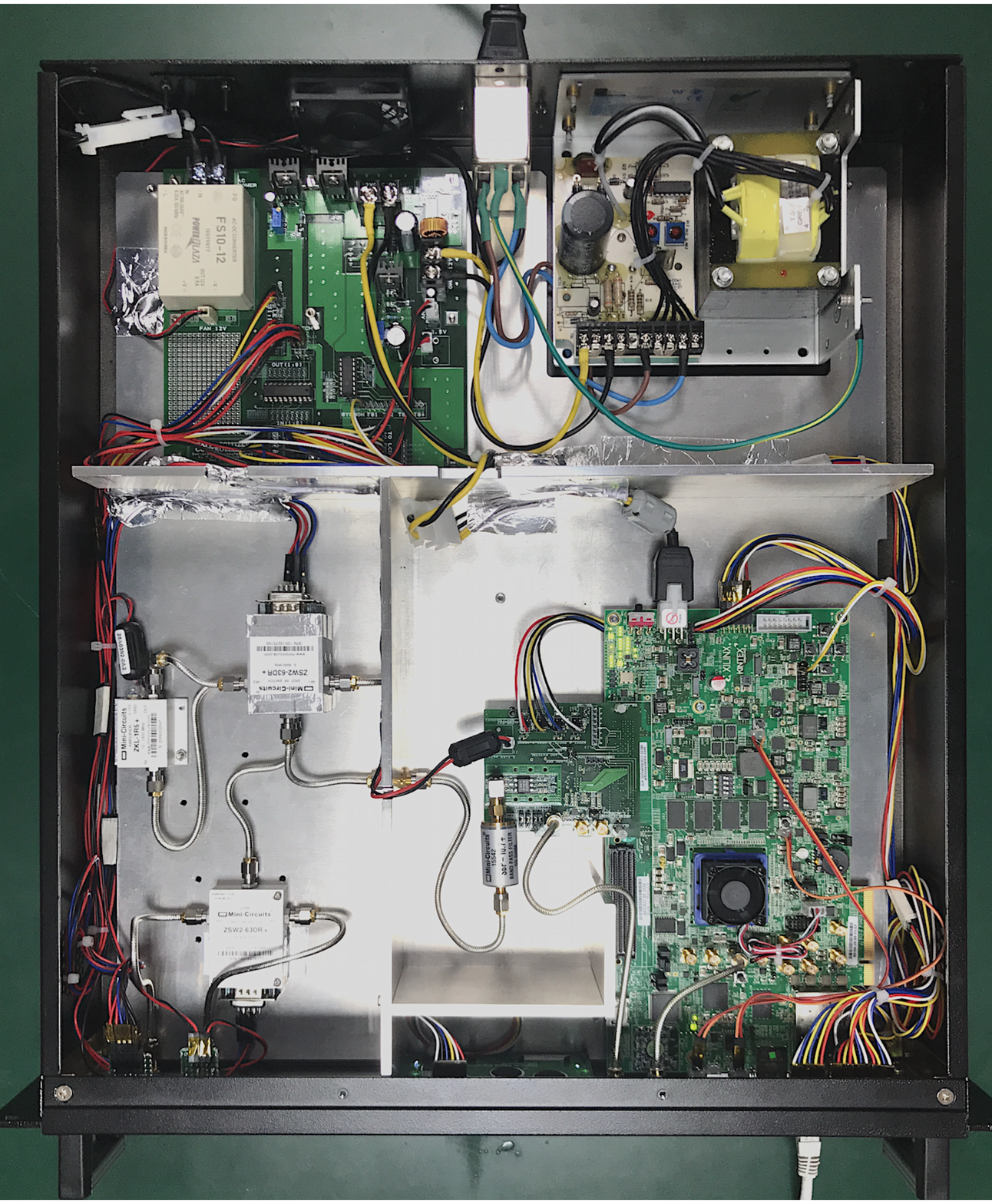}
\caption{A photo of the FPGA DAQ system.  The upper part, lower left part, and lower right part correspond to 
(1) power supply and peripheral interface, 
(2) IF processing path, 
and (3) frontend and FPGA board, respectively. }
\label{fig:digitizer_photo}
\end{center}
\end{figure}

The fabricated   FPGA DAQ system is shown in  Fig. \ref{fig:digitizer_photo}. 
The system is divided into three sections of (1) Power supply and peripheral interface, 
(2) IF processing path,
and (3) frontend and FPGA board. 
The power supply and peripheral interface part includes a linear power supply unit followed by  a few voltage regulators, plus 
several logic translators to drive external components. 
The IF processing path,
the frontend and FPGA board are described in the following subsections. 

\subsection{IF processing path design} 
\label{sec:hardware:rfpath}
\begin{figure}
\begin{center}
\includegraphics[width=0.9\textwidth]{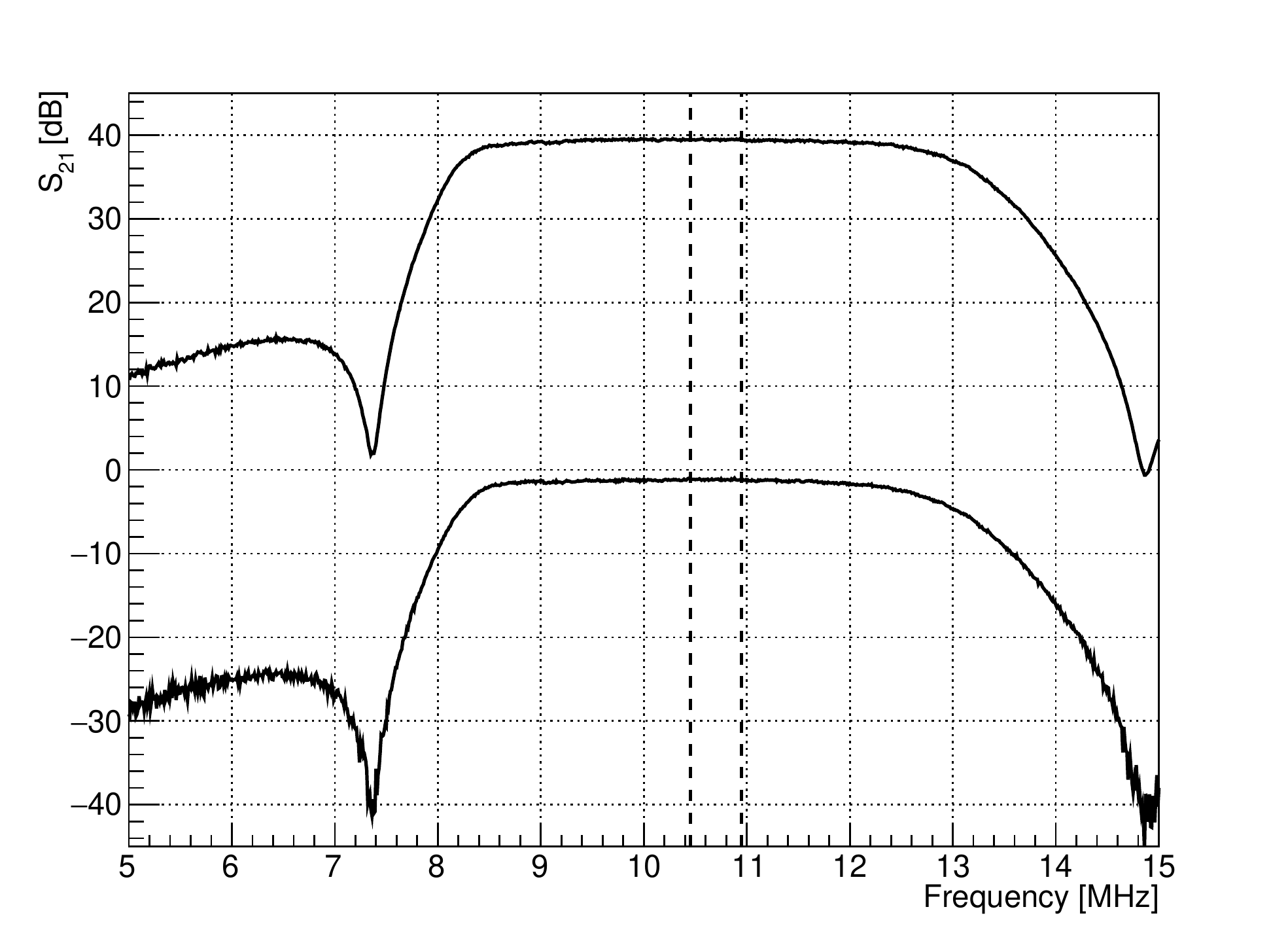}
\caption{Transmission performance of the the 
IF processing path. 
Upper and lower traces 
correspond to using or by-passing an amplifier, respectively. 
The dashed lines correspond to the edges of the analysis window at 10.45 and 10.95\,MHz. 
The gain in the  amplifier and the transmission of the RF switches are almost flat 
in this frequency range, and the characteristic shapes of the traces
mostly reflect the transmission of the band-pass filter. }
\label{fig:rfpath}
\end{center}
\end{figure}

While the FPGA DAQ system can perform FFT 
on the signals from a single channel, it can accept two
different sources of input, such as an input for a calibration signal. 
An RF switch (Mini-Circuits\registered
ZSW2-63DR+, insertion loss 0.33\,dB at 10.7\,MHz) selects the input source. 
Two of the same switch components are used to select (by-) passing the RF amplifier,  for 
additional signal amplification. 
The gain, noise level, and frequency range of the amplifier (Mini-Circuits\registered  ZKL-1R5+) are
40\,dB (typical), 3\,dB (typical), and 10 - 1500\,MHz, respectively. 
The last stage of the 
IF processing path
is equipped with a band-pass filter (Mini-Circuits\registered SBP-10.7+) with a
1.5\,dB pass band of  9.5 - 11.5\,MHz with 0.86\,dB insertion loss at 10.7\,MHz, 
in order to suppress the out-band noise.  
Figure \ref{fig:rfpath} shows the transmission performance of the 
IF processing path 
for both cases, of using amplifier or not. 
In the analysis window, the gains (or losses) of the 
IF processing path
are 39.5\,dB or -1.16\,dB when  using an amplifier or by-passing it, respectively. 

\subsection{ADC and Frontend board}
\label{sec:hardware:adc}
The frontend board is composed of an ADC chip, impedance matching network, clock manipulator, 
and ADC peripheral control circuit. 

The employed ADC is a Texas Instrument\registered  ADS4149\,\cite{ref:WEB_ADS4149} with 14\,bit 250\,MS/s, 
and 2\,V dynamic range. 
It is operated by an external reference clock at 104.8576\,MHz, manipulated through a fast differential logic buffer.  
The sampling clock was chosen considering that the FFT algorithm can be optimized for $2^n$ samples 
($n$ is a positive integer). To achieve 100 Hz frequency width, 
time series data during  $1 / (100\,{\rm Hz}) = 10\,{\rm msec}$ is required. 
Also, from the Nyquist theorem, the sampling rate is supposed to be greater than $10.7 \times 2 = 21.4\,\rm{MHz}$,
and over-sampling is a common practice to decrease the ADC quantization noise. Therefore, 
a sampling rate above 100\,MHz was requested, 
resulting in a  $2^{20} \times 100\,{\rm Hz} = 104.8576\,{\rm MHz}$ sampling rate. 
The sampling clock signal is supplied from  external equipment to get the benefit of a  
high performance signal generator synchronized with a standard clock reference.

Because it is not necessary to change the ADC parameters during operation, 
external ADC control lines are all removed to avoid digital noise from the control signals 
feeding into the RF signal.
The frontend board is connected to the FPGA board through a 400-pin LPC-FMC (FPGA Mezzanine Card) connector. 

\begin{figure}
\begin{center}
\includegraphics[width=0.9\textwidth]{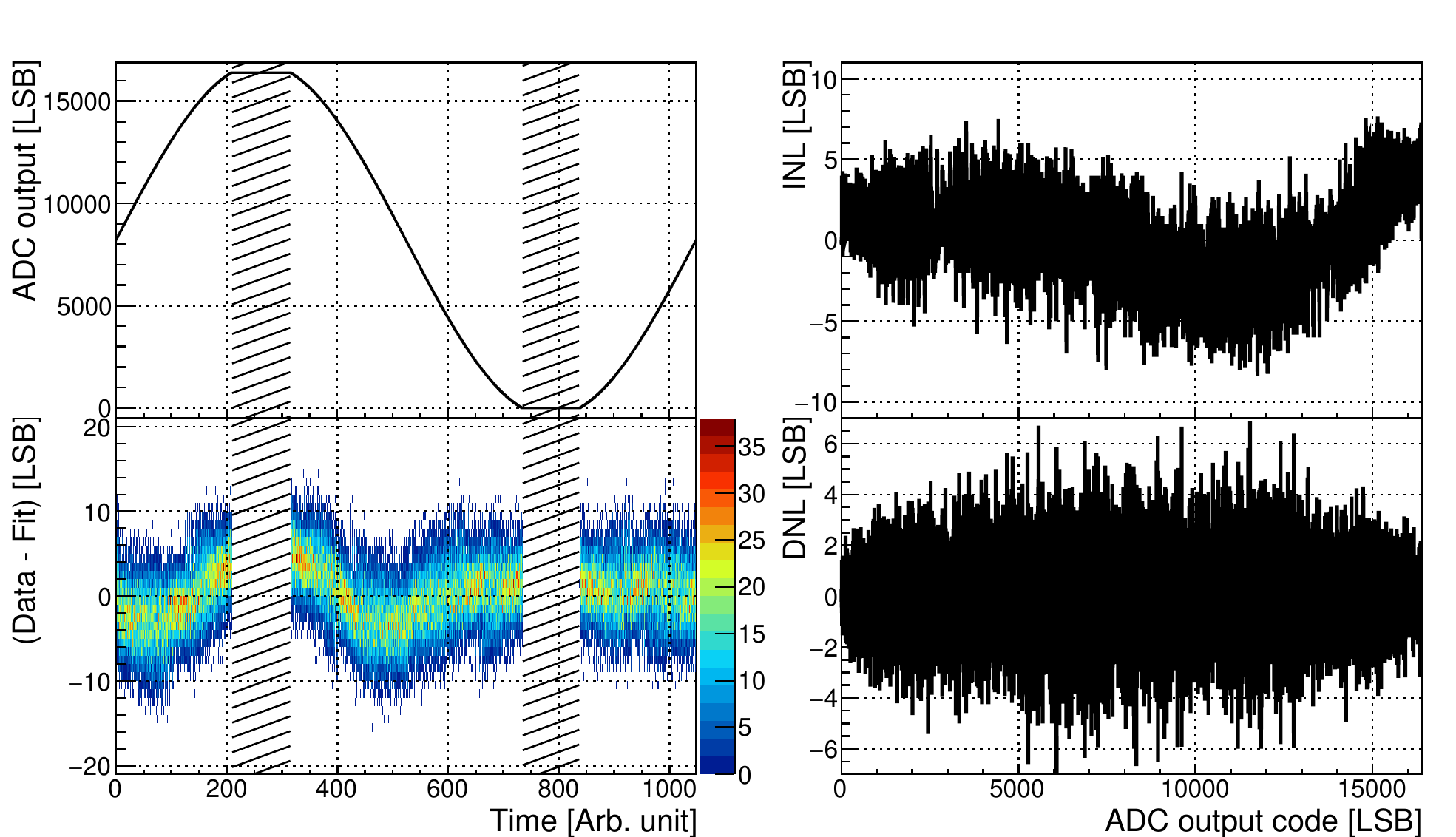}
\caption{The performance of ADC. The left two plots show the ADC output codes in one cycle (top) 
and its fit residual to a truncated sinusoidal function of all (160 cycles) data sample (bottom). 
Note that the hatched area corresponds to the intentional overflow of ADC codes, 
which is not used in the fit residual calculation.  
The right two plots shows INL (top) and DNL (bottom).  }
\label{fig:adctest}
\end{center}
\end{figure}

Figure \ref{fig:adctest} shows the key performances of the ADC, tested with a sinusoidal wave signal. 
The amplitude of the test signal was set slightly bigger than the dynamic range of the ADC, 
in order to avoid any effect near the edge of the dynamic range, 
or distortion of the signal near the local maximum and minimum. 
This results truncate  the observed ADC data, as shown in the left-top plot in Fig. \ref{fig:adctest}. 
A fit of the data to a truncated sinusoidal function was performed 
to estimate the input signal, 
where the fit residual of data in 160  cycles is shown in the left-bottom plot. 
The Integral Non-Linearity (INL) defined by the deviation of the observed ADC code 
from the expected ADC code (from fit), and the
Differential Non-Linearity (DNL) defined by the deviation of the observed ADC codes 
of a bin to the adjacent bin, are shown in the right plot. 
The fit deviation, INL, and DNL are all a few LSB (Least Significant Bit) 
level, which is significantly smaller than the ADC dynamic range (14\,bits). 

\subsection{FPGA board}
\label{sec:hardware:fpga}
For design simplicity, a commercial FPGA evaluation board, 
Xilinx\registered KCU105\,\cite{ref:WEB_KCU105} was used. This board features the
Xilinx Ultrascale\registered XCKU040-2FFVA1156E FPGA.
The FPGA and the evaluation board were chosen considering 
the foreseen logic cell requirement of 
the future logic upgrade of, for example,  multiple analog channel processing.
The board connects the frontend board to receive
digitized 
IF signals, 
and several peripheral components for control and monitoring are connected through General Purpose
Input and Output (GPIO) connectors.

\section{Design of the FPGA firmware}
\label{sec:firmware}
The FFT of $2^{20}$ samples is not practically  possible with FPGA because of limited logic resources. 
However, the width of the signal window is only 500\,kHz, therefore ``Zoomed-FFT'' techniques can be applied:
the time series data is digitally down-converted to baseband. After  applying the digital filter, 
the signal is again sub-sampled for every $2^6$ samples, as the required frequency width is only 100\,Hz. 
The FFT is applied to the sub-sampled data, resulting $2^{14}$ FFT data. 
The lower 5001 points data are sent to the DAQ as a result. 

\begin{figure}
\begin{center}
\includegraphics[width=0.9\textwidth]{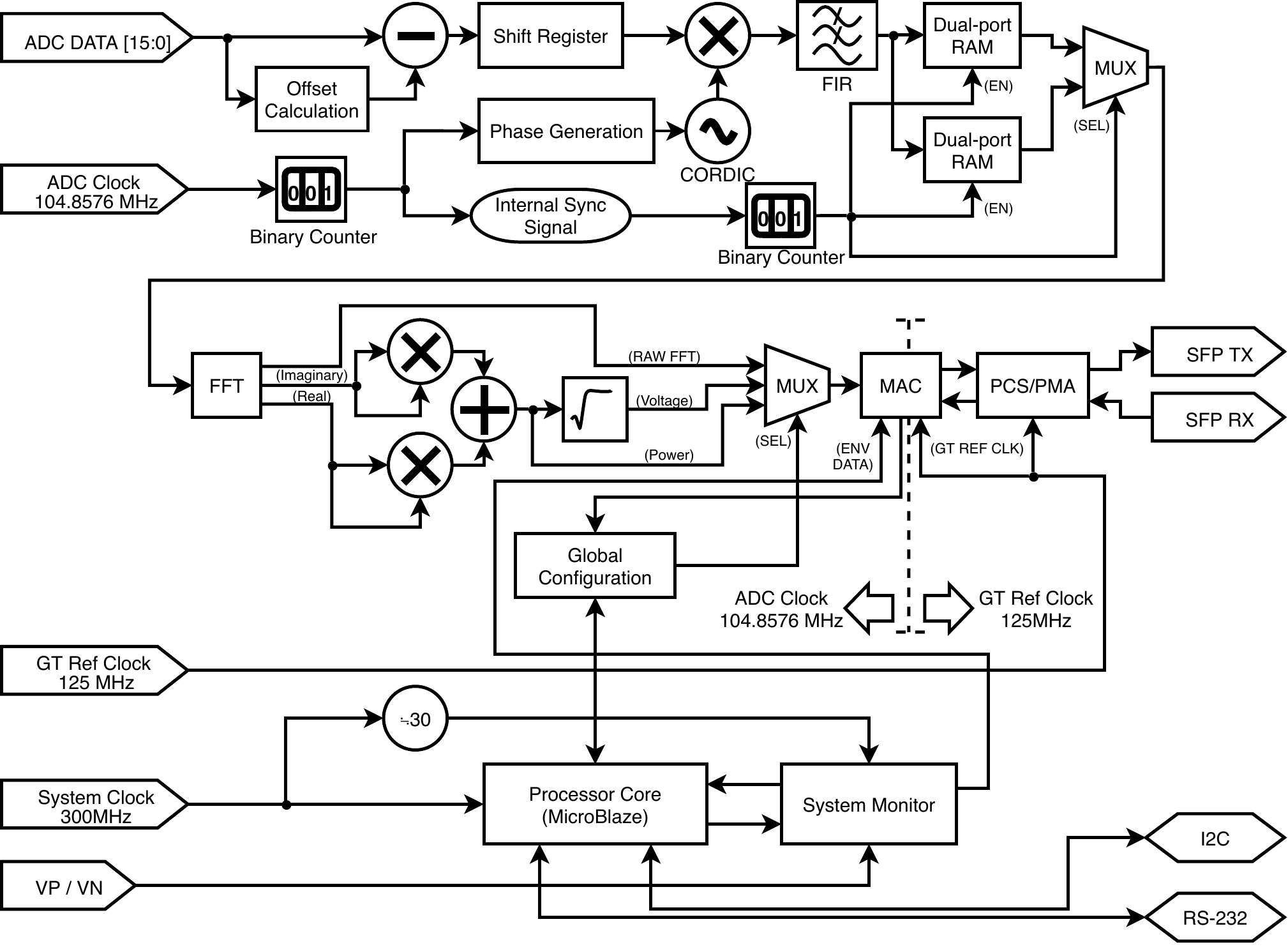}
\caption{The block diagram of the FPGA firmware design of the FPGA DAQ system. }
\label{fig:block_diagram}
\end{center}
\end{figure}

A block diagram of the FPGA firmware is shown in Fig.\,\ref{fig:block_diagram}. 
The core components for the FFT process are explained in the following subsections. 
Those components are operating with the ADC clock at 104.8576\,MHz, while the 
Gigabit Transceiver (GT) operates with a GT reference clock at 125\,MHz,
which is provided by an on-board  jitter cleaner IC. 
The ``Internal Sync Signal'' is generated by counting the ADC clock up to $2^{20}$, 
which is used to synchronize all the FFT processing components. 
The system configuration is done by an Ethernet communication with a DAQ computer, 
which is described in Section\,\ref{sec:firmware:ethernet}. 
A processor core for the system interface and monitor is designed by using 
MicroBlaze Soft Processor Core IP\,\cite{ref:ublaze}
of Vivado\registered  Design Suite\,\cite{ref:vivado}.
The overall utilization of the FPGA resources of the firmware is around 5.4\,\%.

\subsection{Digital mixing for down-conversion}
\label{sec:firmware:mixer}
The signal mixing means multiplying the RF signal with a reference 
sinusoidal signal (the LO signal), 
in order to convert the frequency range of the signal. 
From the Fourier transform, any signal can be written as 
$x(t) = \frac{1}{2\pi} \int_{-\infty}^{\infty}  X(\omega)  e^{i\omega t} d\omega$, 
where $X(\omega)$ represents the amplitude of the signal component with respect to angular velocity $\omega$. 
Assuming a sinusoidal wave  
$x_{LO} (t) = A\,{\rm cos}\,(\omega_{LO} t) = A ( e^{i \omega_{LO} t} + e^{-i \omega_{LO} t}) /2$,
then,
\begin{equation}
x(t) \times x_{LO} (t) = \frac{A}{2\pi} \int_{-\infty}^{\infty} X(\omega) ( e^{i (\omega + \omega_{LO}) t} + e^{ i (\omega - \omega_{LO}) t} ) d\omega~.
\end{equation}
This implies that, by multiplying the sinusoidal reference signal to any signal, 
the frequency divides into two bands, at 
$\omega + \omega_{LO}$ and $| \omega - \omega_{LO} |$. 
By selecting only the upper or lower part, the signal can be up- or down-
converted, respectively, without losing the  information in the signal. 

The axion signal is also down-converted 
from an RF signal to an IF band,
during the signal processing network 
before reaching the DAQ equipment.
In the FPGA DAQ, it is subsequently  down-converted 
from the IF band 
to the  baseband by using 
a digital multiplication algorithm (i.e., not using a hardware mixer component). 
It should be noted that direct down-conversion to the baseband before reaching the ADC 
is not preferred, due to the existence of 
1/f noise which may degrades the axion signal. Also the wide-band mixer component selection 
is limited on the market,
which forces 
to employ multiple RF mixer components that
possibly degrades the axion signal. Since the IF is fixed to 10.7\,MHz and the analysis window is 500\,kHz, 
the LO frequency of digital mixing is determined to be $10.7\,\rm{MHz} - (500\,\rm{kHz})/2 = 10.45\,\rm{MHz}$,
so that the lower limit of the analysis window corresponds to DC after the down-conversion. 
The LO signal is generated by using CORDIC LogiCORE IP\,\cite{ref:cordic} 
of Vivado\registered  Design Suite.

\subsection{Low pass filter}
\label{sec:firmware:lpf}
\begin{figure}
\begin{center}
\includegraphics[width=0.9\textwidth]{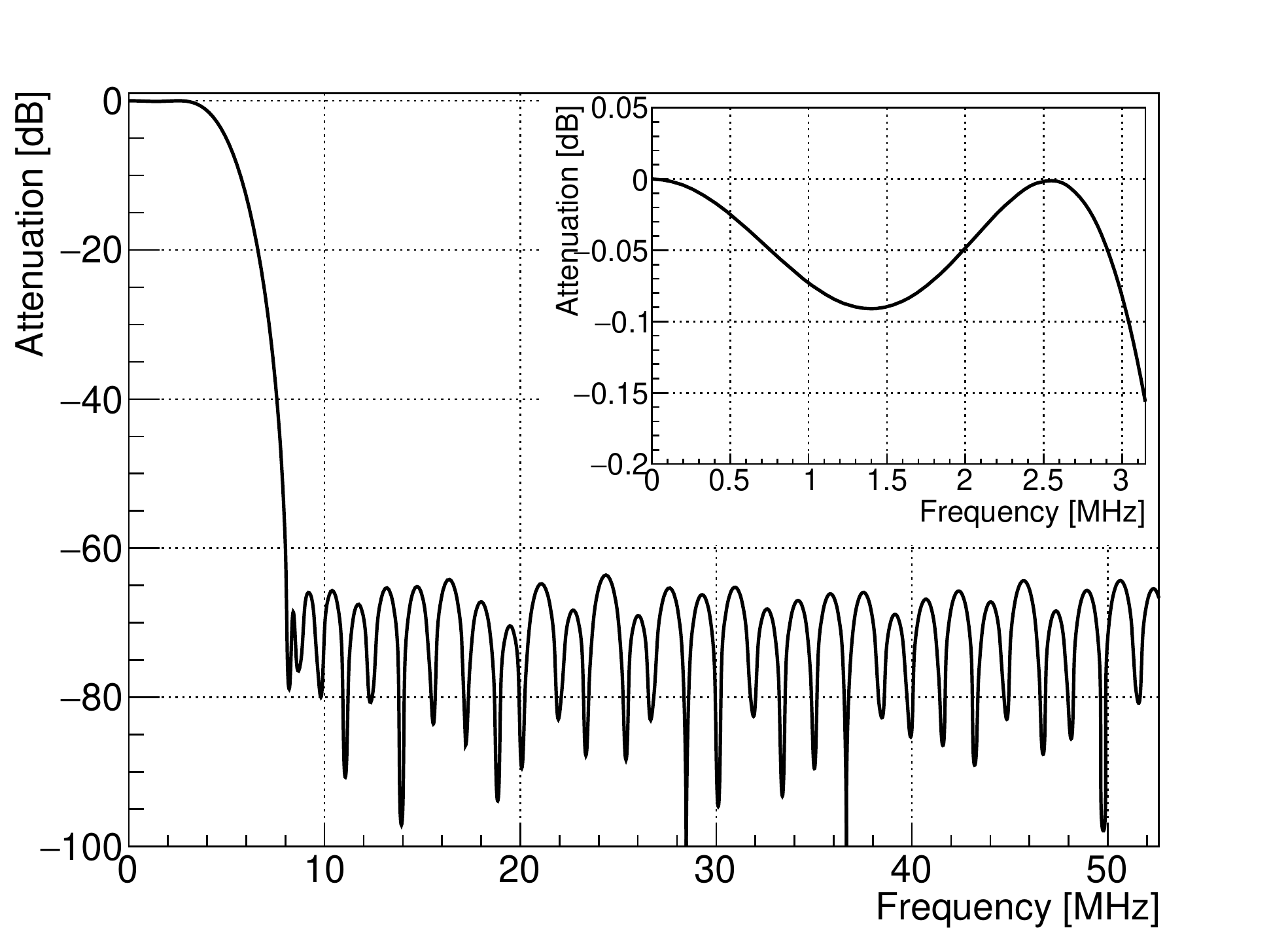}
\caption{The calculated frequency response of the FIR filter. The inset figure shows that of the passband. }
\label{fig:fir}
\end{center}
\end{figure}

Since the 
IF signal 
will be mixed to two side band signals, 
corresponding to the sum and difference of the 
IF
and LO frequencies, 
the low pass filter is equipped to select the lower-band signal only. 
The Finite Impulse Response (FIR) filter is designed using the 
Parks-McClellan optimal FIR filter design\,\cite{ref:octavefir} of GNU Octave program\,\cite{ref:octave},
while it is implemented in the FPGA logic using the
FIR Compiler LogiCORE IP\,\cite{ref:fir} of Vivado\registered Design Suite. 
The analysis window after down-conversion corresponds to DC - 500\,kHz, 
and the high-side image signal is located around 21.15\,MHz, 
therefore, the passband of FIR filter is designed up to 3\,MHz with an attenuation ripple less than 0.1\,dB. 
The stopband of the FIR is defined from 8\,MHz, with attenuation of at least -60\,dB. 
The calculated frequency response is shown in Fig. \ref{fig:fir}. 

\subsection{Sub-sampling and FFT}
\label{sec:firmware:fft}
The maximum frequency of the baseband signal is 500\,kHz. 
For efficient subsampling of $2^{20}$ samples while maintaining a 
frequency width of 100\,Hz, the subsampling rate of $2^{14} \times 100\,\rm{Hz} = 1.6384\,\rm{MHz}$  is applied,
resulting in a 1/64 subsampling to 16384 ($=2^{14}$) samples out of $2^{20}$ samples. 
This subsampling rate is more than two times that of the analysis window (500\,kHz), 
satisfying the Nyquist theorem. 
To achieve real-time and pipelined processing, dual block RAMs were constructed in the FPGA logic, 
in order to 
record next 16384 samples while  transacting the previous  16384 samples to the FFT logic.

The FFT process was designed using the 
Fast Fourier Transform LogiCORE IP\,\cite{ref:fft} of Vivado\registered Design Suite. 
It calculates 16384 points of FFT data, but only the lower 5001 points are selected (up to 500\,kHz) 
and sent to the DAQ. 
The architecture of the FFT is the Radix-2 Decimation In Frequency (DIF) method (``Pipelined  Streaming IO''). 
The FFT calculates the real and imaginary parts simultaneously, to yield signal power in a linear scale 
in an arbitrary unit. 

\subsection{Ethernet protocol design} 
\label{sec:firmware:ethernet}
To enable real-time data transaction to the DAQ, a special Ethernet protocol was designed in the FPGA logic. 
The data rate of the FFT can be calculated as follows. The size of a point in the FFT result is 8\,bytes,
and 5001 data points are calculated during 10\,msec. Then, ignoring the transaction overhead, 
the averaged data rate is at least $8 \times 5001 / 0.01 = 4\,\rm{Mbyte/sec}$ or 32\,Mbps. 
In order to provide sufficient transaction bandwidth,
1\,GbE Ethernet was employed to connect the system to the DAQ. 
The merit of using 1\,GbE is that interface hardware 
and software with very high reliability are commonly available, 
from which the system design becomes significantly simpler. 
Also, it allows any future design upgrades with higher data rates. 

For simplicity in design, one-to-one direct connection between the FPGA DAQ system and DAQ computer without any network routing device in between is assumed. 
The protocol or packet structure for FPGA DAQ is designed based on a custom, non-standard Ethernet 
(in the data link layer (layer-2) of the OSI model\,\cite{ref:osimodel}) packet design. 
The entire packet structure is shown in Fig. \ref{fig:packet_structure}. 
The fields related to the FPGA DAQ are explained below. 
Other fields are defined in IEEE 802.3\,\cite{ref:ieee802}. 
\begin{description}
\item[EVTID] This indicates the event number of the FPGA DAQ data, which is supposed to increase by one at the start of the new sampling every 10\,msec. 
\item[PKTID] Since one event's data is split into several Ethernet packets, this field indicates the packet number inside the same event. 
32 packets are required to send 5001 FFT data point. 
\item[TIMESTAMP] This 32\,bits-long  field contains the clock data for  the time of packet being sent, with $10\,\mu\rm{sec}$ resolution. 
The field value is used to check whether the packets are being sent in real-time or any transaction error has occurred. This field completes the FPGA DAQ data header. 
\item[SAMPLE DATA] This 160-times recurring field  contains  FFT data. Every Ethernet packet sends 160 samples, therefore, 1280 bytes in total. 
\item[IDENTIFIER] This field value is fixed to 0x4341505000 (``CAPP\textbackslash0'') in order to identify the end of the data stream. 
\item[MODE] This field contains the DAQ mode information. 
\item[NSAMPLE] The number of FPGA DAQ data points in this packet. If every transaction is alright, this value should be 160 for all samples up to the 31-th packet, 
and 41 for the last (32nd) packet. This field completes the FPGA DAQ data trailer, and also completes the Ethernet payload. 
\end{description}

A Media Access Controller (MAC) is customary designed for handling this packet structure and converting it into Gigabit Media Independent Interface (GMII). 
The Physical Coding Sublayer (PCS) and Physical Medium Attachment (PMA) receive the GMII data and send it to the DAQ computer through the 
1\,GbE link. The PCS/PMA was designed using 1G/2.5G Ethernet PCS/PMA or SGMII LogiCORE IP\,\cite{ref:pcspma} of the Vivado\registered Design Suite.

\begin{figure}
\centering
\begin{tabular}{l|p{9mm}|p{9mm}|p{9mm}|p{9mm}|p{9mm}|p{9mm}|p{9mm}|p{9mm}|}
 Byte & 1 & 2 & 3 & 4 & 5 & 6 & 7 & 8 \\
\hline
 & \multicolumn{7}{|c|}{PREAMBLE} & \multicolumn{1}{|c|}{SFD} \\
\cline{2-9}
 Ethernet header & \multicolumn{6}{|c|}{DESTMAC} & \multicolumn{2}{|c|}{SRCMAC}\\
\cline{2-9}
 & \multicolumn{4}{|c|}{SRCMAC} & \multicolumn{2}{|c|}{TYPE} & \multicolumn{2}{c}{} \\
\hline
 FPGA DAQ header& \multicolumn{3}{|c|}{EVTID} & \multicolumn{1}{|c|}{PKTID} & \multicolumn{4}{|c|}{TIMESTAMP} \\
\hline
 & \multicolumn{8}{|c|}{SAMPLE DATA \#1} \\
\cline{2-9}
 FPGA DAQ data & \multicolumn{8}{|c|}{$\cdot \cdot \cdot$} \\
\cline{2-9}
 & \multicolumn{8}{|c|}{SAMPLE DATA \#160} \\
\hline
FPGA DAQ trailer & \multicolumn{5}{|c|}{IDENTIFIER}  & \multicolumn{1}{|c|}{MODE} & \multicolumn{2}{|c|}{NSAMPLE} \\
\hline
Ethernet trailer & \multicolumn{4}{|c|}{FCS} & \multicolumn{4}{c}{}\\
\cline{1-5}

\end{tabular}
\caption{The entire Ethernet packet structure of the FPGA DAQ data. See the text for details. }
\label{fig:packet_structure}
\end{figure}

\section{Performance of the FPGA DAQ system}
\label{sec:test}

The first conceptual design of the FPGA DAQ was prepared  in IBS/CAPP and fabricated by an external company along with the FPGA logic design in 2018. 
The design was revised in 2019 to improve the IF processing path and frontend circuit to reach a better noise level.
The revised products were fabricated and tested in IBS/CAPP during 2020 and deployed in one of the  CULTASK experiment for  further tests in a real experimental environment. 
The test results on the noise level, linearity of signal measurement, and DAQ efficiency are described below.

\subsection{Noise level}
\label{sec:test:noise}
\begin{figure}
\begin{center}
\includegraphics[width=0.9\textwidth]{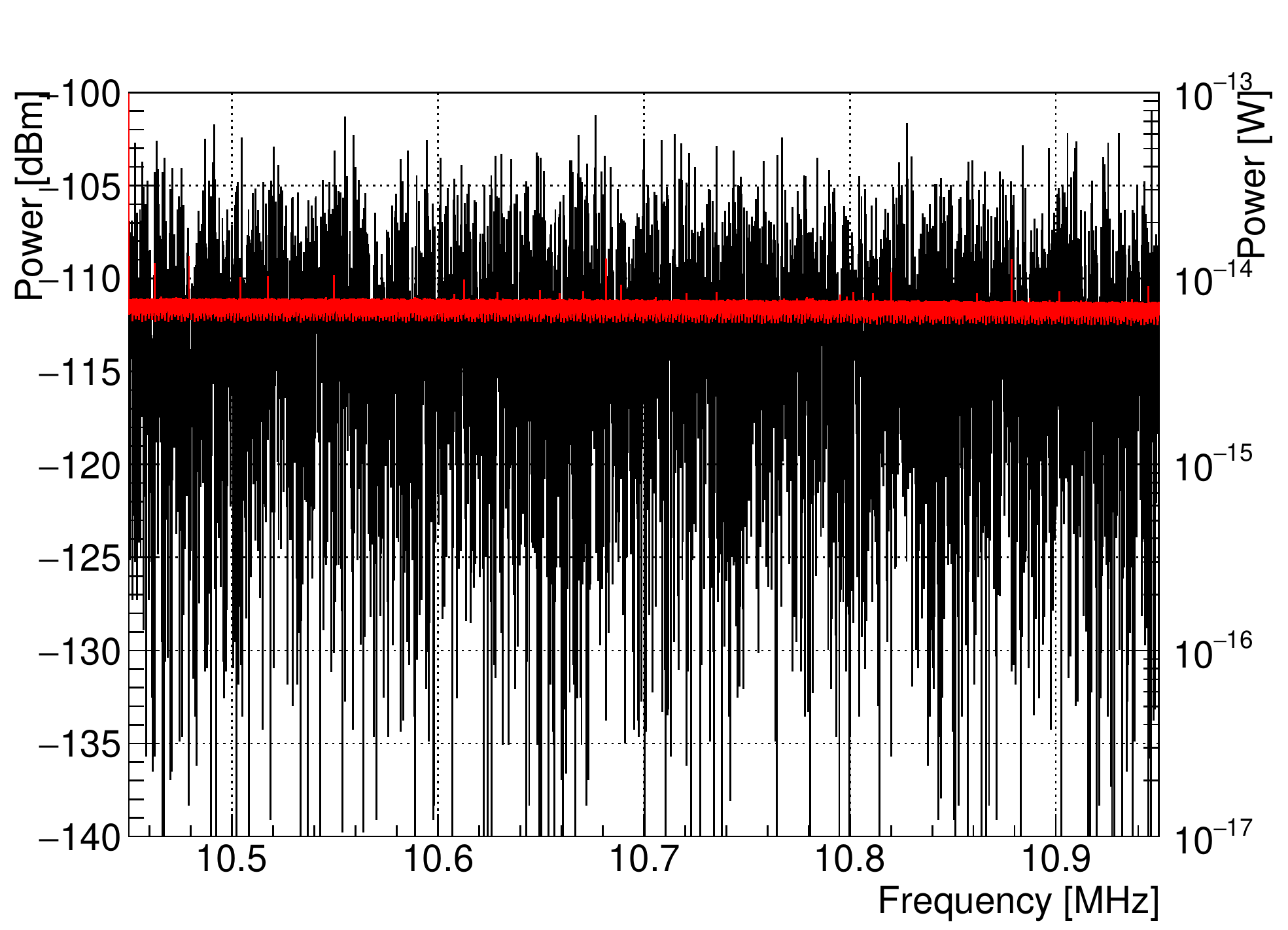}
\caption{The noise level spectrum of a single measurement (black) and 100,000 times average (red). }
\label{fig:bg_spectrum}
\end{center}
\end{figure}

Figure \ref{fig:bg_spectrum} shows the power spectrum with a 50\,$\Omega$ terminated input, 
corresponding to the effective noise level of
the FPGA DAQ system, after applying the calibration result discussed in 
Section\,\ref{sec:test:calibration}.
The noise level of a single measurement (the black trace in Fig. \ref{fig:bg_spectrum}) 
fluctuates along the frequency, 
ranging from around -150\,dBm to -100\,dBm. Because the noise voltage is Gaussian-distributed, and the power of
each spectrum bin follows $\chi^2$ distribution of degree 2, averaging  the  number of measurements in a 
spectrum bin reduces the fluctuation in the measurement, following the radiometer equation, as discussed in detail in 
Section\,\ref{sec:test:pseudodata}.
The red trace in Fig. \ref{fig:bg_spectrum} shows the average of 100,000 times measurements, clearly showing the
reduction in noise fluctuation. The averaged noise level of 100,000 times measurements is -111.7\,dBm, and 
the effective noise floor defined by the maximum of the noise distribution is -108.8\,dBm. 
The first spectrum bin somehow shows an incorrect measurement almost every time, which can be ignored in the real measurement. 
It also shows clear evidences of spurious peaks which might come from the signal mixing or an external noise intrusion, 
which, however, does not affect the axion experimental measurement 
of which the noise level after  amplification through the RF receiver chain is much higher than the spurious peaks.

\subsection{Frequency reconstruction and gain stability along frequency}
\label{sec:test:gain}
\begin{figure}
\begin{center}
\includegraphics[width=0.9\textwidth]{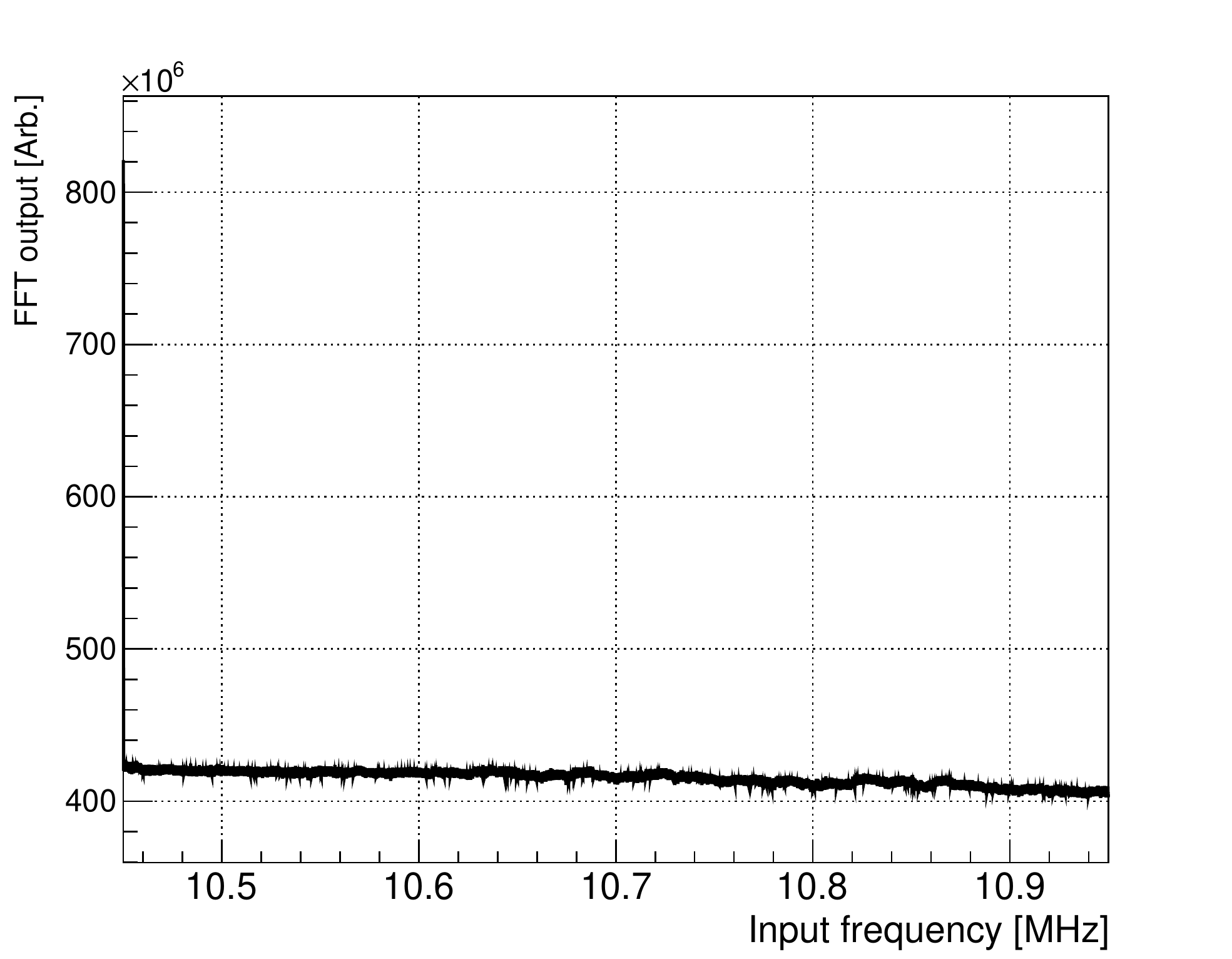}
\caption{
The measured power of the input signal along the frequency when injecting -60\,dBm signal, representing the gain stability along the frequency. 
The power calibration is not applied, 
therefore, the vertical axis is in an arbitrary unit.
}
\label{fig:freq_linearity}
\end{center}
\end{figure}

Correct reconstruction of the measurement frequency is very important in the axion experiment,
as the frequency represents the axion mass. 
While the measurement frequency is controlled by the tuning system of the cavity resonant frequency 
and the RF signal down-conversion to IF before reaching to the FPGA DAQ, 
the FPGA DAQ have to measure the frequency correctly without any distortion. 
In order to confirm this, 
the dependence of the measured frequency bin number (defined as $i_{\rm meas}$) to the input frequency, 
and the existence of missing $i_{\rm meas}$, were examined. 
The input signal was  $-60\,\rm{dBm}$ sinusoidal signal with the frequency ranging 
from 10.45\,MHz to 10.95\,MHz
with 100\,Hz step,
and the peak position (representing $i_{\rm meas}$) and its height of FFT (i.e., measured power of input signal before applying calibration)   were measured. 
The $i_{\rm meas}$ ranges from 0 to 5000, supposedly corresponding to 10.45\,MHz and 10.95\,MHz. 
The input signal frequency was converted to the expected input frequency bin number, $i_{\rm input}$, by
\begin{equation}
i_{\rm input} = (({\rm Input~frequency})- 10.45\,{\rm MHz}) / 100\,{\rm Hz}~. 
\label{eq:freqrecon}
\end{equation}
Over the entire frequency range from 10.45\,MHz to 10.95\,MHz, the differences between $i_{\rm meas}$ and  $i_{\rm input}$  
were zero,
meaning that the signal frequency can be calculated using Eq.\,\ref{eq:freqrecon} without any discrepancy. 
All $i_{\rm meas}$ were evenly distributed over the entire frequency range without any missing value in the measurement window. 
The measured power distribution 
over the entire frequency range is shown in Fig. \ref{fig:freq_linearity}. 
A notable gain drop of a maximum 0.1\,dB was observed when the frequency was increasing, partly 
due to the variation in gain  of
the low pass filter, as shown in Fig. \ref{fig:fir}. 
The measured  power of the first frequency bin (corresponding 10.45\,MHz) was always comparably 
different than the flat gain distribution,
which is not understood yet. 
This does not affect  the real experiment, since 
the corresponding frequency data can be 
discarded every time as it is away from the cavity resonant frequency. 
This discarding  will result in reducing the measured frequency range 100 Hz only.

\subsection{Linearity of power measurement and power calibration}
\label{sec:test:calibration}
\begin{figure}
\begin{center}
\includegraphics[width=0.9\textwidth]{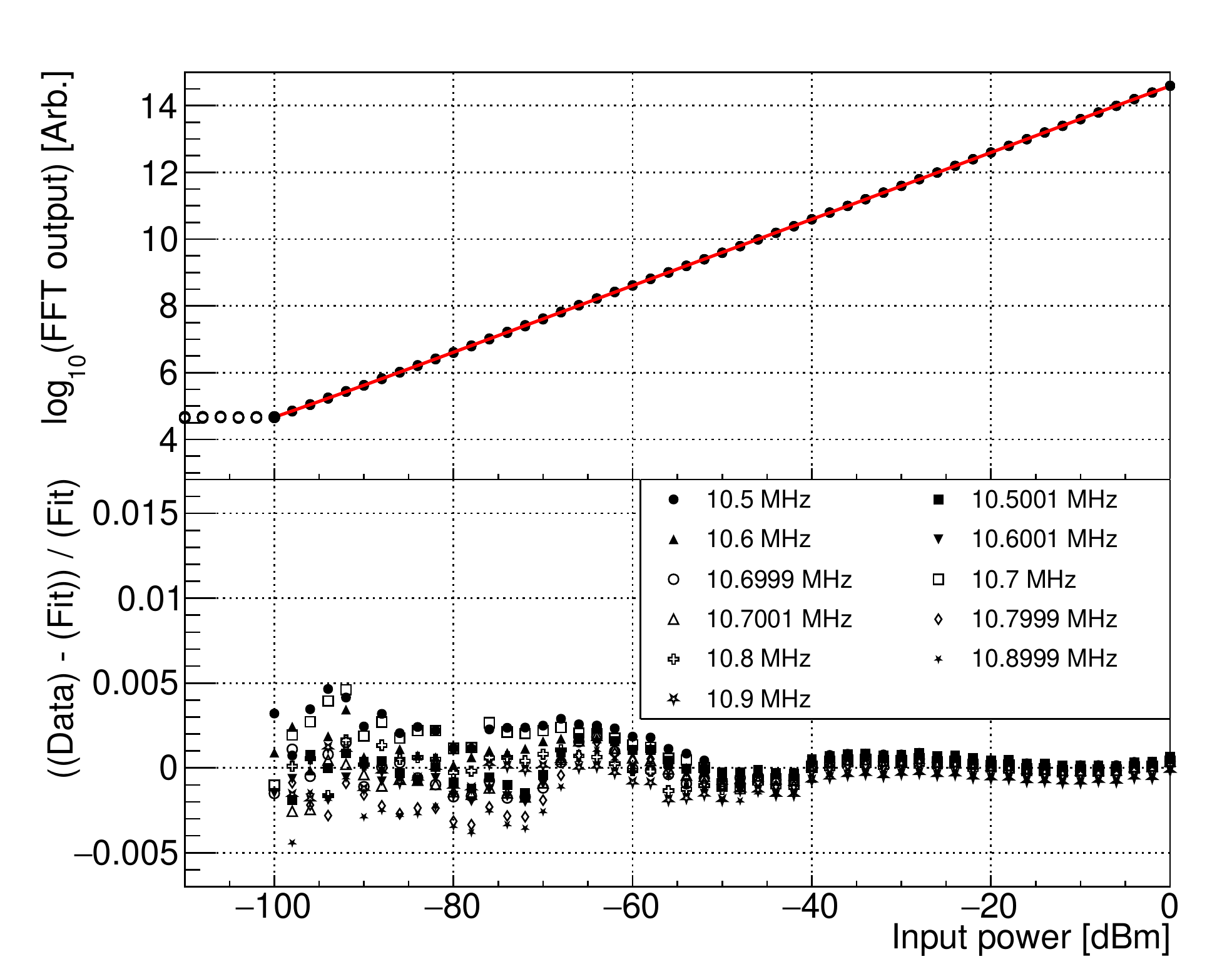}
\caption{The power calibration data. 
The upper plot includes all the measurement data in the filled and hollow points, representing
correct frequency measurements of the signal and incorrect frequency measurements (meaning that a signal was not observed), respectively. 
The red solid line shows the fit to a 5th-order polynomial function. 
The bottom plot shows the relative deviation of the data from the calibration model  to demonstrate the
quality of the fit, with different marker types corresponding to different measurement frequencies. }
\label{fig:power_linearity}
\end{center}
\end{figure}

While the frequency measurement is exact in the FFT output, the power measurement requires calibration 
using an external reference source. 
The calibration procedure uses a signal generator for the  calibrated power, and the FFT output codes are measured
while  varying power. Several different frequency points are measured  to understand the frequency dependence. 
The data with the correct signal frequency recovery (the filled points in the upper plot 
in Fig. \ref{fig:power_linearity}) are used for 5th-order polynomial function fit to obtain the calibration model. 
From the relative deviation of the model, shown in the lower plot of Fig. \ref{fig:power_linearity},
the uncertainty from the calibration was estimated to be less than 0.5\,\%, in the worst case 
of near-noise floor measurement.
A clear evidence of the dependence to the frequency was observed, 
for example, the relative deviation were always positive for 10.5\,MHz data, while always negative for 10.8999\,MHz data. 
This is partly because of the gain change along the frequency as described in the Section\,\ref{sec:test:gain}. 
The relative deviation  also shows a fluctuation between the nearby frequency bins, of which the reason is not understood yet. 
However, all these dependence are less than the overall calibration uncertainty of 0.5\,\%. 
The calibration dependence on the temperatures of various parts of the FPGA DAQ 
and the bias voltages were checked, however, no significant dependence was observed 
in an one-day continuous DAQ data. 
A long term stability is still being checked with a CULTASK experiment setup.

\subsection{DAQ efficiency}
\label{sec:test:efficiency}
\begin{figure}
\begin{center}
\includegraphics[width=0.9\textwidth]{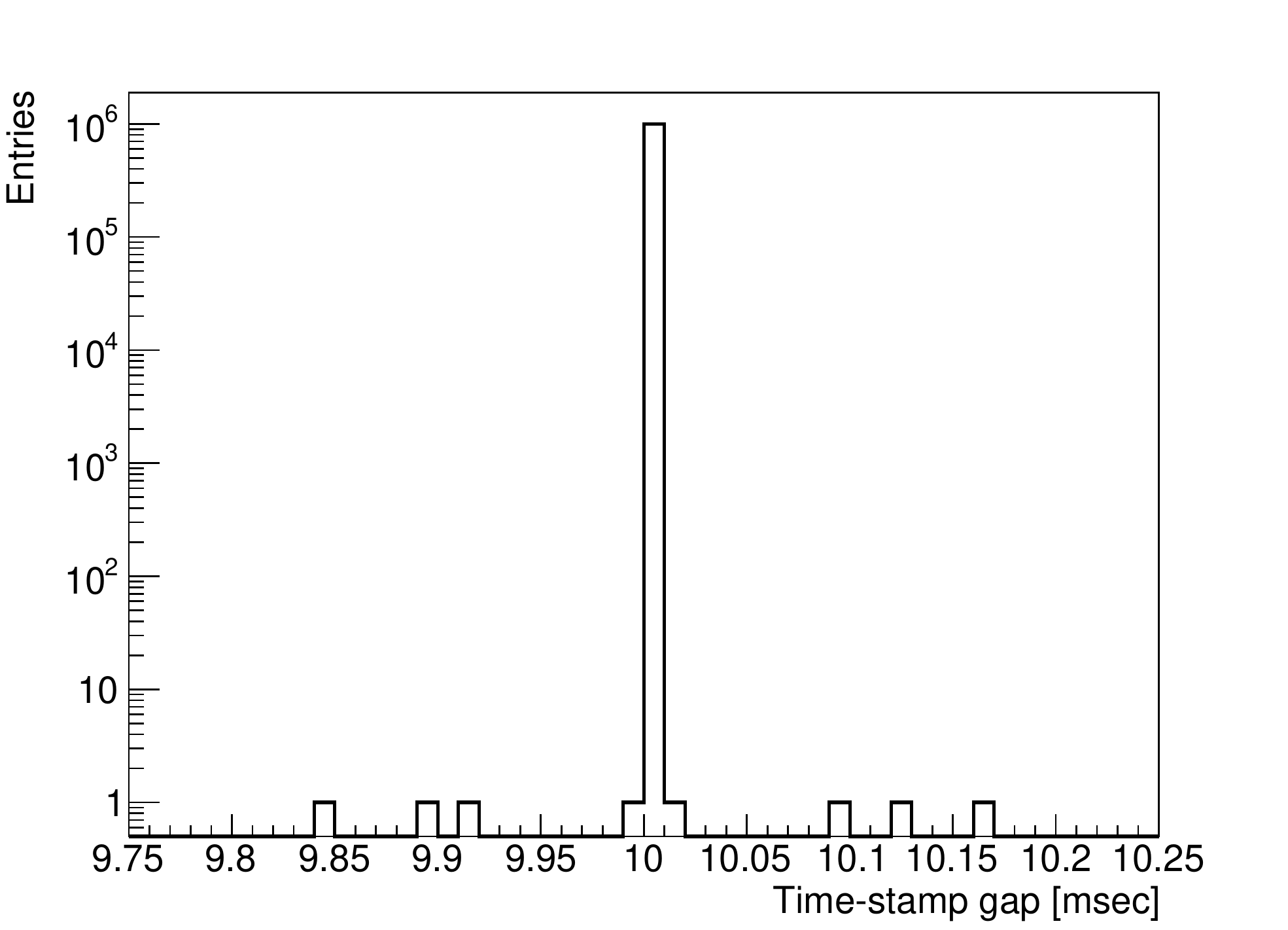}
\caption{The distribution of time-stamp differences between adjacent events. }
\label{fig:tsgap}
\end{center}
\end{figure}

The DAQ efficiency was estimated by using the time-stamp data of the data packet. Every Ethernet packet contains 
time-stamp data indicating the $10 \mu \rm{sec}$-resolution wall time when the packet is sent to the DAQ computer. 
Figure \ref{fig:tsgap} shows the distribution in time-stamp differences between adjacent events, 
for $1\times 10^{6}$ events from 2.8 hours continuous DAQ. 
The network interface card effectively filters out any events with an incorrect Frame Check Sequence (FCS), 
and the event number recorded in this data sample increased by one for all events, 
which means no data loss occurred during the
Ethernet 
transmission.
The observed time-stamp difference between adjacent events was 
$10.000 \pm 0.001\,{\rm msec}$ 
in average, 
without any overflowing or underflowing events. 
This means all the FFT data 
during this measurement time
was completely transferred and recorded in the DAQ. 

It should be noted that this 100\,\% DAQ efficiency could be achieved not only owing to the special design on the pipelined signal processing 
and  the Ethernet packet and protocol of the FPGA DAQ, but also due to the idle status of the DAQ computer. 
The Network Interface Card  and its driver software of the DAQ computer may not contain 
enough  buffer memory in order not to discard any data that was not read by the operating system. 
Therefore, in order not to lose any data transmitted from the FPGA DAQ, the DAQ computer is expected to remain idle status. 
This requires a dedicated DAQ computer for the FPGA DAQ, not shared by any other part of DAQ.

\subsection{Axion signal detection analysis with pseudo-data}
\label{sec:test:pseudodata}
Assuming the  system temperature of the cryogenic system of axion experiment is 1\,K, 
the noise power is $k_B T_{\rm sys} \Delta\nu = 1.38 \times 10^{-21}\,\rm{W}$ where 
$k_B$ is the Boltzmann constant, $T_{\rm sys}$ is the system temperature, and $\Delta \nu$ is frequency resolution.
The postulated axion signal power is around $10^{-22} - 10^{-23}\,\rm{W}$, which is lower than the noise level. 
The Signal-to-Noise Ratio (SNR) of the axion detection 
can be improved by increasing the number of  measurements\,
\cite{ref:admx, ref:haystac},
as depicted in the Radiometer equation:
\begin{equation}
{\rm SNR} = \frac{T_{\rm src}}{T_{\rm sys} / \sqrt{N}} = \frac{P_{\rm src}}{k_B T_{\rm sys}} \sqrt{\frac{\tau}{\Delta \nu}}~~,
\label{eq:radiometer}
\end{equation}
where $P_{\rm src} = k_B T_{\rm src} \Delta \nu $ is the axion signal power of equivalent temperature $T_{\rm src}$, 
and $N = \tau \Delta \nu$ is the number of measurements during the measurement time $\tau$. 
As a demonstration of the axion signal detection principle using the Radiometer equation, 
in this pseudo-data test, a combined signal of arbitrary white noise and a sinusoidal signal of a known frequency 
and various power levels 
were supplied to the FPGA DAQ system,  to understand the 
SNR improvement over the repeated measurement. 

\begin{figure}
\begin{center}
\includegraphics[width=0.9\textwidth]{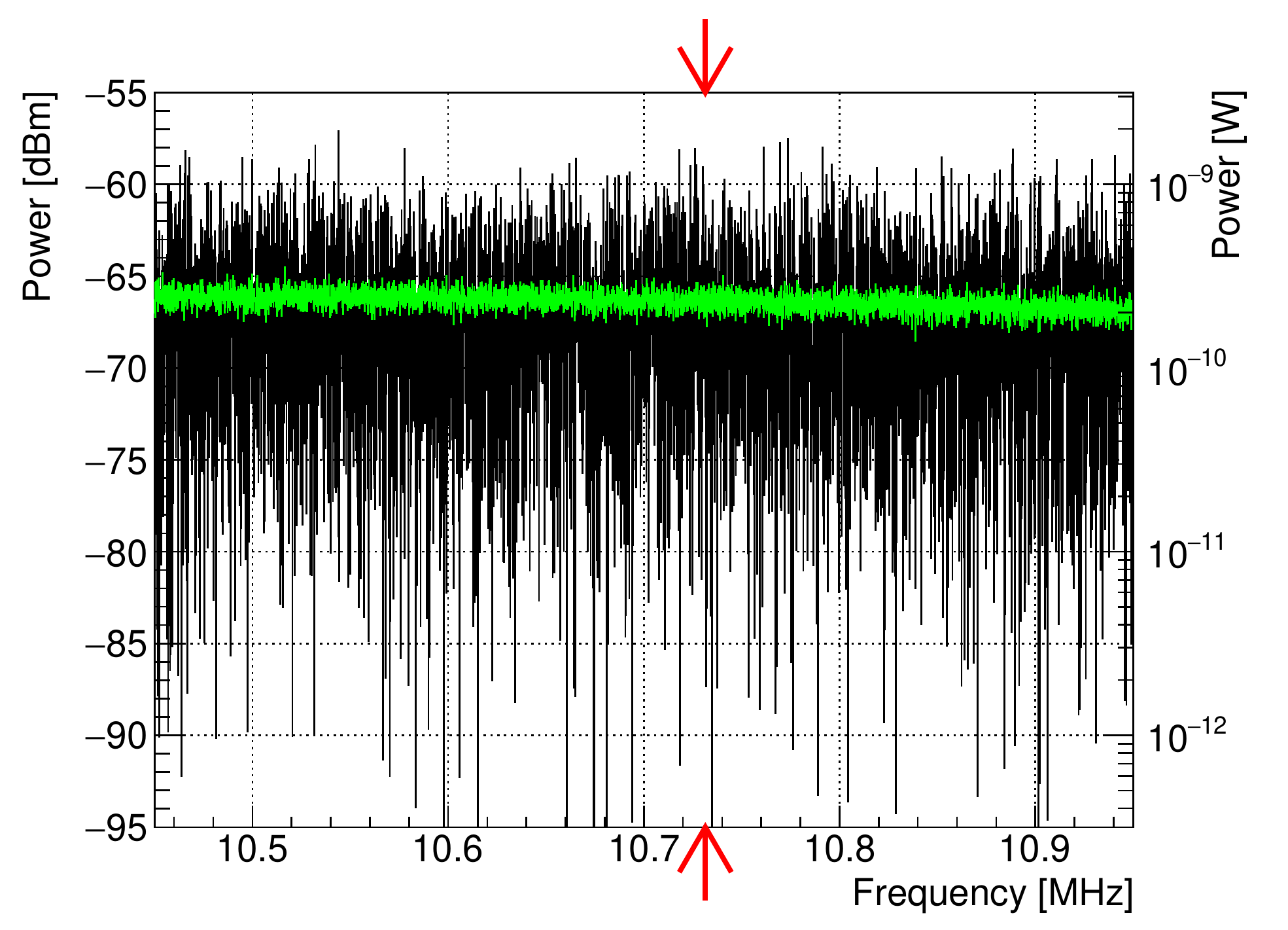}
\caption{The pseudo axion experiment data of a single measurement (black) and 100 times average (green). The red arrow represents the frequency location of the -83.8\,dBm signal at 10.73141592\,MHz.  }
\label{fig:noisesig_bgdist}
\end{center}
\end{figure}

The measured noise level of the CULTASK experiment 
is around $-60 \sim -40$\,dBm after passing through several cryogenic 
and room-temperature amplifiers. The white noise supplied to the FPGA DAQ system was set to around -66.3\,dBm, and 
generated by an arbitrary function generator. 
Another RF signal generator was used to generate 10.73141592\,MHz signal which was combined with the white noise
through an RF combiner component. 
Figure \ref{fig:noisesig_bgdist} compares  the measured spectrum of a single measurement (black) 
and 100 times average (green), when the input signal power is -83.8\,dBm combined with -66.3\,dBm white noise.
In this measurement, the signal is not observed. 

\begin{figure}
\begin{center}
\includegraphics[width=0.9\textwidth]{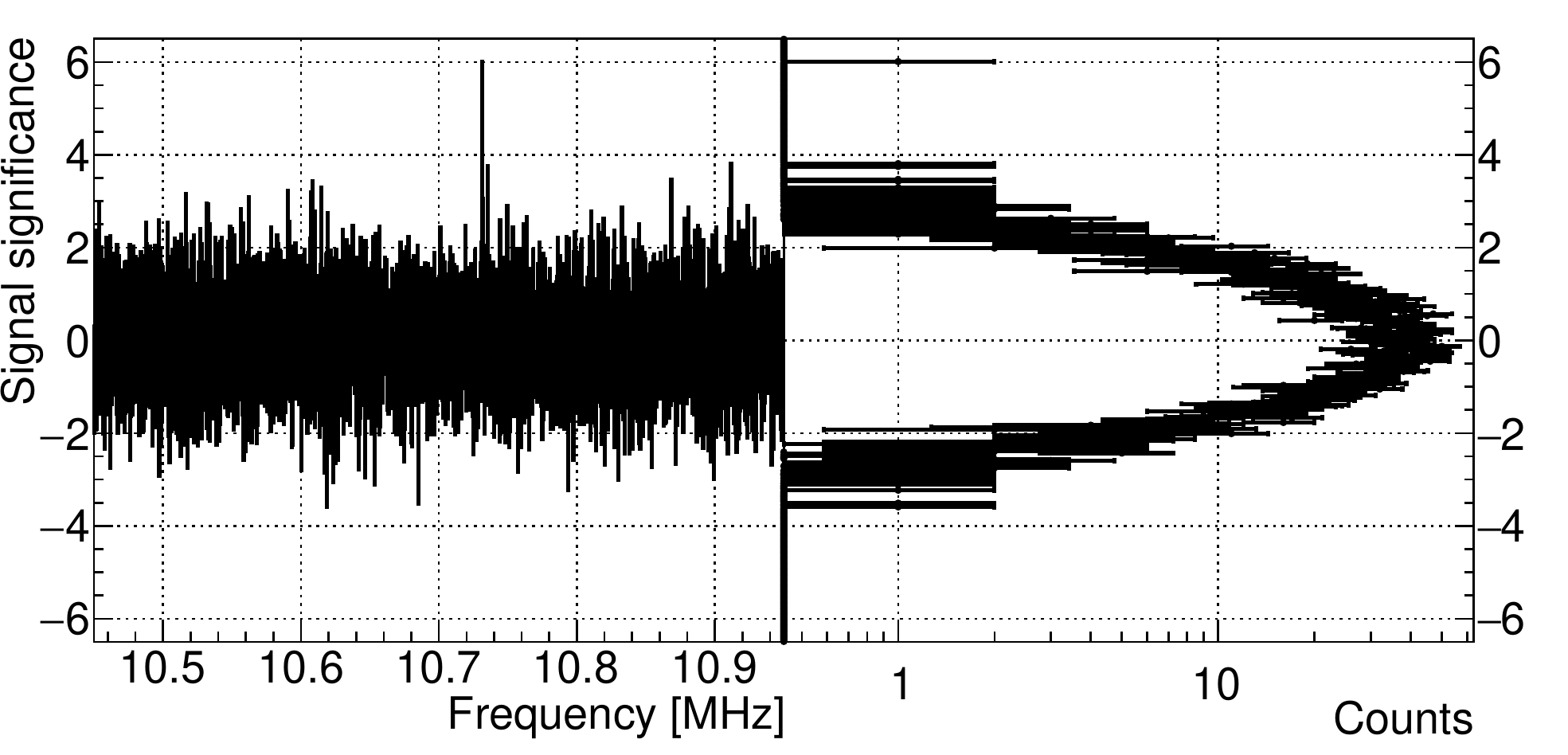}
\caption{The standard normal deviation spectrum (left) and histogram (right) of the signal significance of 100,000 times measurement of the pseudo-axion experiment, with the -83.8\,dBm signal combined with -66.3\,dBm white noise. }
\label{fig:noisesig_pull}
\end{center}
\end{figure}

The pseudo-axion experimental spectrum was measured 100,000 times and averaged.
In order to determine  the power excess, the  averaged spectrum was divided by the baseline estimation from the
Savitzky-Golay filter, and one is subtracted\,\cite{ref:haystac}.
The standard normal deviation of the power excess histogram, regarded as  the signal significance 
or the SNR, is shown in Fig. \ref{fig:noisesig_pull}. 
In this figure, the signal power was  -83.8\,dBm, while the average number  was 100,000. 
The signal at 10.7314\,MHz was clearly observed with $\sim6.0\sigma$ deviation. 
Note that the SNR estimated from the radiometer equation was 5.6, which is comparable with this measurement. 

\begin{figure}
\begin{center}
\includegraphics[width=0.9\textwidth]{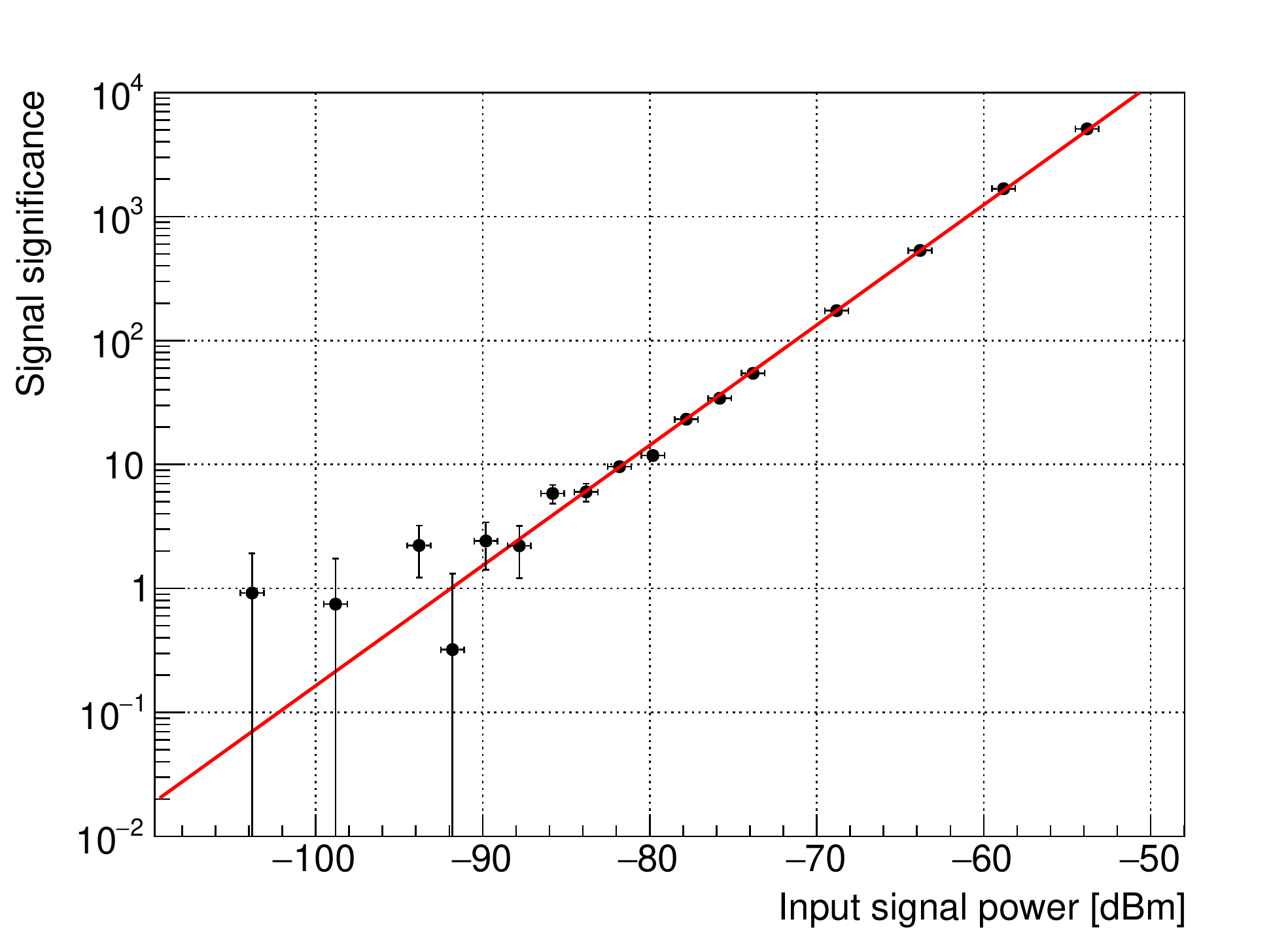}
\caption{The signal significance dependence to the input signal power with 100,000 times average and -68.3\,dBm noise level. The red curve represents a fit to an exponential function. From the fit, the  $3\sigma$ signal significance corresponds to -87.0\,dBm. }
\label{fig:noisesig_significance}
\end{center}
\end{figure}

Assuming the $3\sigma$ deviation represents the possible existence of an axion signal candidate,
the input signal power was scanned from near -50\,dBm to -100\,dBm, with 100,000 times average for each scan. 
The signal significance was calculated for each power scan, and the result is shown in Fig. \ref{fig:noisesig_significance}. 
From an exponential function fit, it was estimated that a -87.0\,dBm signal will result in a 
$3\sigma$ signal significance under -66.3\,dBm white noise, when 100,000 spectra are averaged. 
In a real experiment, frequency points of more than $3\sigma$ significance is regarded as an axion signal
candidate frequency, which is subject to a re-scanning experiment for further SNR improvement. 

\section{Summary and conclusion}
\label{sec:summary}
An FPGA based real-time DAQ system was constructed and tested 
for the CULTASK axion experiment. 
The FPGA DAQ system is optimally designed for RF spectrum measurement in the axion experiment, 
especially with regard to DAQ efficiency. 
The 100,000 times averaged noise level of the FPGA DAQ system was around -111.7\,dBm which meets the requirement. 
It was confirmed that the repeated measurement would  increase the SNR, following
the radiometer equation. In case of 100,000 times measurement, under -68.3\,dBm noise level, 
the axion signal of -87.0\,dBm can be detected with $3\sigma$ significance. 
Several FPGA DAQ systems are constructed in IBS/CAPP and will be used as the main DAQ equipment of the CULTASK experiment.

\section*{Acknowledgments}

This work was supported by the Institute for Basic Science (IBS-R017-D1-2021-a00) of Korea. 
The author M. J. Lee would like to thank Mr. Byung-gyu Kim and others of Ubicom technology, Korea, for their help on constructing
the first version of the FPGA DAQ system.

\end{document}